\documentstyle[prd,aps,floats]{revtex} 
\input{epsf}

\def\AJ{{\it Ap. J.} }
\def\AJL{{\it Ap. J. Lett.} }

\def\CQG{{\it Class. Quantum Gravity} }

\def\FP{{\it Fortschr. Physik} }
\def\GRG{{\it Gen. Relativity and Gravitation} }

\def\JHEP{{\it JHEP} }

\def\MPL{{\it Mod. Phys. Lett.} }
\def\MNRAS{{\it Mon. Not. R. Ast. Soc.} }
\def\NAT{{\it Nature} }
 
\def\NC{{\it Il Nuovo Cimento} }

\def\PL{{\it Phys. Lett.} }
\def\PR{{\it Phys. Rev.} }
\def\PRL{{\it Phys. Rev. Lett.} }

\def\be{\beta}

\def\frac#1#2{{\textstyle{{#1}\over {#2}}}}

\def\lsim{\mathrel{\rlap{\lower4pt\hbox{\hskip1pt$\sim$}}
    \raise1pt\hbox{$<$}}}
\def\gsim{\mathrel{\rlap{\lower4pt\hbox{\hskip1pt$\sim$}}
    \raise1pt\hbox{$>$}}}
\def\sqr#1#2{{\vcenter{\vbox{\hrule height.#2pt
         \hbox{\vrule width.#2pt height#1pt \kern#1pt
         \vrule width.#2pt}
         \hrule height.#2pt}}}}

\def\be{\begin{equation}}
\def\ee{\end{equation}}
\def\bea{\begin{eqnarray}} 
\def\eea{\end{eqnarray}}

\begin{document}

\twocolumn[\hsize\textwidth\columnwidth\hsize\csname
@twocolumnfalse\endcsname

\title{A Two-Field Quintessence Model}

\author{M. C. Bento$^{1,2}$, O. Bertolami$^1$ and N. C. Santos$^1$}

\address{$^1$  Departamento de F\'\i sica, Instituto Superior  T\'ecnico}

\address{$^2$  Centro de F\'{\i}sica das 
Interac\c c\~oes Fundamentais, Instituto Superior  T\'ecnico\\
Av. Rovisco Pais 1, 1049-001 Lisboa, Portugal}

\address{E-mail addresses: bento@sirius.ist.utl.pt; orfeu@cosmos.ist.utl.pt}

\vskip 0.5cm

\date{\today}

\maketitle

\begin{abstract}
We study the dynamics of a quintessence model based on two interacting
scalar  fields. The model can account for the (recent) accelerated
expansion of the Universe suggested by astronomical
observations. Acceleration can be permanent or temporary and, for both 
scenarios, it is possible 
to obtain suitable values for the cosmological parameters while 
satisfying the nucleosynthesis constraint on the quintessence energy
density. We argue that the model dynamics can be made consistent with a stable 
zero-energy relaxing supersymmetric vacuum.

\vskip 0.5cm
 
\end{abstract}

\pacs{PACS numbers: 98.80.Cq,95.35+d \hspace{2cm}Preprint DF/IST-7.2001, 
FISIST/09-2001/CFIF}

\vskip 2pc]

Recent  observations  of  type  Ia supernovae,  together  with  Cosmic
Microwave  Background  (CMB) and  cluster   mass  distribution  data
\cite{Perlmutter,Boomerang,Bahcall} indicate that the Universe is
flat, in agreement with  the inflationary prediction, accelerating and
that the energy density of (baryonic plus dark) matter 
is smaller than the  critical density. Thus, observations suggest that
the dynamics  of the Universe at present is dominated by  a negative
pressure component, the main candidates being a cosmological constant
and a slowly-varying vacuum energy,  usually
referred to  as dark energy or  quintessence \cite{Ratra,Caldwell} (an
evolving vacuum energy was  discussed somewhat earlier,
e.g. \cite{Bertolami1}). The main difference between the cosmological
constant  and quintessence  scenarios is  that, for  quintessence, the
equation of state parameter,  $w_Q\equiv p/\rho$, varies with time and
approaches a present  value $w_Q<-0.6$,  whilst for the cosmological
constant, it remains fixed at $w_\Lambda=-1$.

Several quintessence models have been put foward, most of them based
on a scalar field which was sub-dominant in the early Universe and,
more recently, has started to dominate the energy density of
non-relativistic matter. Theoretical suggestions include a scalar
field endowed  with exponential \cite{Peebles,Ferreira,Skordis,Barrow}
or inverse power law potentials \cite{Zlatev}, the string theory
dilaton  in the  context of gaugino condensation \cite{Binetruy}, an
axion field  with an almost massless quark \cite{Kim}, scalar-tensor
theories of gravity \cite{Uzan}, or one of the fields arising from the
compactification process in the multidimensional
Einstein-Yang-Mills system \cite{Bento1}. Some of these models
address the ``cosmic coincidence'' problem {\it i.e.} the question
of explaining why the vacuum energy or scalar field dominates the
Universe  only recently. In tracker models, the tracker field
rolls down a potential according to an attractor-like solution to
the equations of motion, causing the energy density of the
quintessence field to track the equation of state of the
background  energy component  independently of  initial conditions
\cite{Zlatev}. 
However, in these  models, the overall scale of the
potential has to be fine-tuned in  order for the quintessence
energy to overtake the matter density at present.
 
In $k$-essence
models \cite{Steink}, as a result of the dynamics, tracking of the
background energy density can only occur in the radiation epoch;
at the onset of matter-domination, the $k$-essence field energy
density drops sharply, increasing again and overtaking the matter
energy  density at roughly the current epoch. At least in the
original proposal, these features require the introduction of a
non-linear kinetic energy density functional of the scalar field and
adjusting it to obtain the desired attractor behaviour.

A commom feature of the  proposals mentioned above is that the asymptotic
accelerating behaviour of the Universe is driven by the dynamics of a
single field. In this  work, we shall consider instead a two-field
model. Two-field quintessence models were previously considered, in an attempt 
to explain how to obtain a small 
but non-vanishing cosmological constant 
\cite{Fujii} and in the context of SUSY QCD \cite{Masiero}. 
Actually, there are several motivations for studying  potentials with
coupled scalar  fields. In fact, if one envisages to extract a
potential suitable for describing the Universe dynamics from
fundamental theories, it is  most likely that an ensemble of scalar
fields (moduli, axions, chiral  superfields, etc) will emerge, for
instance, from the compactification process or from the localization of
fields in the brane in multi-brane models or from  mechanisms
responsible for the cancellation of the cosmological constant (see
e.g. \cite{Burgess} and  references  therein). Furthermore, coupled
scalar fields are invoked for various desirable features they exhibit, as
in the so-called hybrid inflationary models
\cite{Linde,Bento2} and in reheating models in the presence
\cite{Kofman} or absence \cite{Bertolami2} of parametric resonance.
Finally, it has been recently pointed out that an  eternally
accelerating Universe poses a challenge for string theory, at least in
its present formulation, since asymptotic states are inconsistent with
spacetimes that exhibit event horizons \cite{Hellerman}.
Moreover, it is argued that  theories with  a  stable supersymmetric
vacuum cannot relax into a zero-energy ground state if the
accelerating  dynamics is guided by a single scalar field
\cite{Hellerman}. In this paper, we present a
two-field model whose solutions allow for, at least, a partial fixing
of these inconsistencies.

Another interesting feature of our model  
is that it presents  two types  of solutions,
namely one in which the Universe accelerates forever and one in which
it is possible for the Universe to exit from a period of accelerated
expansion and resume decelerated expansion. The latter type of solution is
compatible with the conceptual framework underlying string theory.


It is believed that scalar fields with potentials of the type

\be
V(\phi, \psi)=  e^{-\lambda\phi} P(\phi, \psi)~,
\label{pot}
\ee
where $P(\phi, \psi)$ contains polynomial as well as interacting
terms in $\phi$ and $\psi$, arise in the low-energy
limit of fundamental particle physics theories such as string/M-theory 
\cite{Choi}, 
$N=2$ Supergravity coupled with matter in higher dimensions \cite{Salam} 
and phenomenological brane-world  constructions \cite{Burgess,Goldberger}. 
The overall negative exponential term
in $\phi$ signals that this is a moduli type field which has
acquired an interacting  potential with the $\psi$ field. A simple
possibility is   

\bea  
P(\phi, \psi)&=     &      A~+
(\phi-\phi_0)^2+B~(\psi-\psi_0)^2  \nonumber \\
&& + C~\phi(\psi-\psi_0)^2 + D~\psi (\phi-\phi_0)^2~,
\label{pot1}
\eea
in units where $M \equiv (8\pi G)^{-1/2}=\hbar=c=1$. Notice that
this potential, for $B=C=D=0$, in which case only the $\phi$ field is
present, coincides with the one proposed in Ref.~\cite{Skordis},
hereby referred to as the AS model. We have considered just tree
interacting terms as they  already capture the main aspects of the
coupled dynamics we are interested in. As we will show, an important
property of model is that it can lead to an asymptotic dynamics where
either $\psi$ or both fields do not necessarily settle in their
minima at present, which is the  key to evade some of the conclusions
of Refs. \cite{Hellerman}, concerning the stability of the
supersymmetric vacua.
    
We consider a spatially-flat  Friedmann-Robertson-Walker  (FRW)
Universe containing a perfect  fluid with barotropic equation of state
$p_\gamma=(\gamma - 1) \rho_\gamma$,  where $\gamma$ is a constant, $0
\leq \gamma \leq 2$ (for radiation $\gamma=4/3$ and for dust $\gamma =
1$)   and  two  coupled   scalar  fields   with  potential   given  by
Eq.~(\ref{pot}).   The evolution  equations for  a  spatially-flat FRW
model with Hubble parameter $H\equiv \dot a /a$ are 
\bea
\dot{H} & = &
- \,   {1\over2}   \left(   \rho_\gamma   +  p_\gamma   +   \dot\phi^2
+\dot\psi^2\right)      \      ,      \\
\dot\rho_\gamma      &=& -3H(\rho_\gamma+p_\gamma)     \    ,     \\  
\ddot\phi    &=& -3H\dot\phi-{\partial_\phi  V}\  ,\\
\ddot\psi  &=&  -  3H\dot\psi  -{\partial_\psi V}\ ,
\label{fried}
\eea
where $\partial_{\phi(\psi)} V  \equiv {\partial V \over \partial
\phi(\psi)}$, subject to the Friedmann constraint

\be  
H^2  =  {1\over3}   \left(  \rho_\gamma   +  {1\over2}\dot\phi^2
+{1\over2} \dot\psi^2 +V \right) \ ,
\label{Friedmann}
\ee
The total energy density of the homogeneous scalar fields is given
by $\rho_Q=\dot\phi^2/2+\dot\psi^2/2+ V(\phi, \psi)$.

A necessary and sufficient condition for the Universe to accelerate is
that the deceleration parameter, q, given by

\be
q = - {a \ddot a\over {\dot a}^2}={1\over 2}  (1+3 w_Q \Omega_Q +
\Omega_r)~,
\label{q}
\ee
where $\Omega_r$ is the fractional radiation energy density, is negative.


We now study how the solutions of the system above depend on
the parameters of the  potential and initial conditions. We  integrate from 
N=-30, corresponding to the Planck epoch; nucleosynthesis occurs around
N=-10, radiation to matter transition around N=-4 and N=0 today.

\begin{figure}[t]
\centering
\leavevmode \epsfysize=10cm \epsfbox{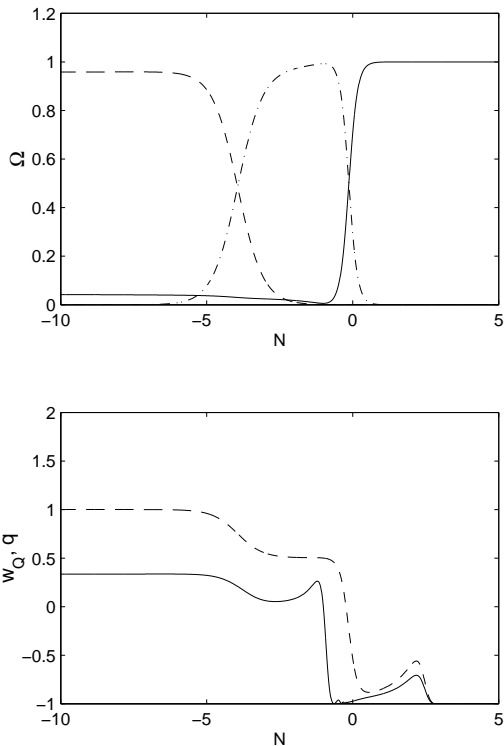}\\
\caption{The upper panel shows the evolution of $\Omega_Q$ (solid),
$\Omega_{r}$ (dashed) and $\Omega_m$ (dot-dashed) for Model I, 
corresponding  to  permanent
vacuum domination. The lower panel shows the evolution of  $w_Q$ 
(solid) and  $q$ (dashed).}
\label{fig:model1}
\end{figure}

\begin{figure}[t]
\centering 
\leavevmode\epsfysize=10cm \epsfbox{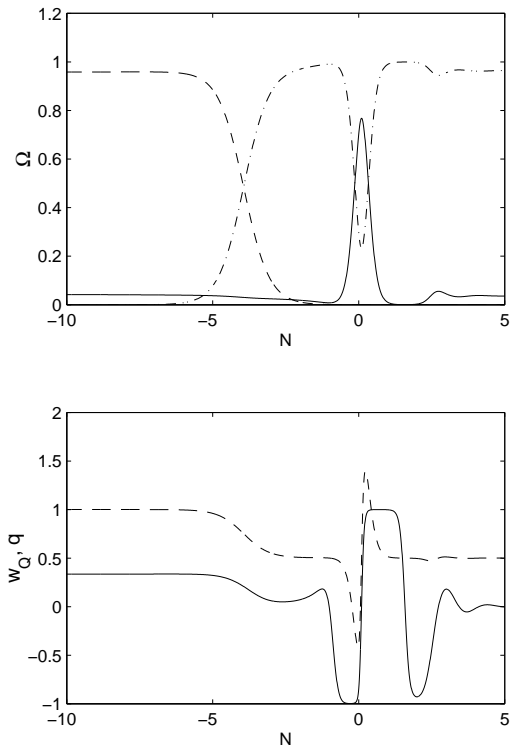}\\
\caption{The upper panel shows the evolution of $\Omega_Q$ (solid),
$\Omega_{r}$ (dashed) 
and  $\Omega_m$ (dot-dashed) for Model II, corresponding  to temporary vacuum
domination. The lower panel shows the evolution of $w_Q$ (solid) and $q$ 
(dashed).}
\label{fig:model2}
\end{figure}

There are essentially two realistic types of behaviour, illustrated in
Figures~\ref{fig:model1}  (Model I)  and \ref{fig:model2}  (Model II).
Model I $(\lambda=9.5, A=0.02, \phi_0=29, \psi_0=15$,  $B=0.002, 
C = 6 \times 10^{-4}, D=4.5)$  corresponds to the  case where
vacuum domination,  which occurs when  $\Omega_{Q0}>1/2$, is permanent
and Model II $(\lambda=9.5, A=0.1, \phi_0 = 29, \psi_0 = 20$, $B =
0.001, C =8\times 10^{-5}, D = 2.8)$ to  the case where vacuum
domination is transient. Permanent  and transient vacuum domination
have  also been found in the AS model, in  Refs.~\cite{Skordis} and
\cite{Barrow}, respectively. Of course, there remains the
(non-realistic) case where accelerated expansion never occurs.

In both models the equation of state has reached $w_Q\simeq -1$ for
the present time, as favored by the available data \cite{Efstathiou} 
(and making it hard to distinguish from a cosmological constant) but,
whereas in Model I $\omega_Q$  remains negative, in Model II it is
in the process of increasing today towards positive values, 
then oscillates slightly  until it reaches its 
asymptotic value. Similarly, in both models, the deceleration parameter
is negative today  but, whereas for Model I 
$q$  remains negative, in Model II it oscillates and becomes positive before
it reaches its  asymptotic value.
   
\begin{figure}[t]
\centering
 \leavevmode\epsfysize=7cm \epsfbox{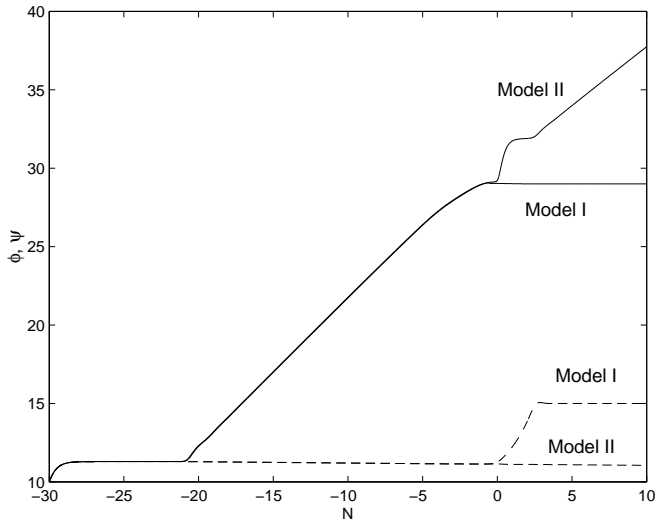}\\
\caption{Evolution  of  the quintessence  fields,  $\phi$ (solid)  and
$\psi$ (dashed), for Models I and II. }
\label{fig:fields3}
\end{figure}

Permanent vacuum domination takes place when at least the $\phi$ field
ends up settling  at the minimum of the  potential, thus corresponding
to a cosmological constant;  in  Model I, both fields settle at the 
minimum of the potential, see Figure~\ref{fig:fields3}. Transient 
vacuum domination occurs either
when the potential has no local minimum or $\phi$ arrives at the local
minimum with enough kinetic energy to roll over the barrier and resume
descending the potential. Notice that the
evolution of $\psi$ is slight compared with $\phi$, 
especially for Model II.

Both models satisfy present bounds on relevant cosmological observables.   
The tightest  bound  comes from  nucleosynthesis, 
$\Omega_Q(N\sim -10) <  0.044$,
requiring $\lambda>9$ \cite{Bean}. The bound  arising
from the most recent CMB data, $\Omega_{Q}  < 0.39$ at last scattering, 
is less stringent than the nucleosynthesis bound. Other
bounds we take into account  are: $\Omega_{m} = 0.3 \pm 0.1$, $w_{Q}
< - 0.6$, $\Omega_{Q} = 0.65 \pm 0.05$ \cite{Perlmutter1} and
$h=0.65  \pm   0.05$  \cite{Bahcall} today. Indeed, Model I has $h=0.6$,
$\Omega_{Q} = 0.7$, $\Omega_{m} = 0.3$, $w_{Q}=-1$ and $q=-0.5$
today, $\Omega_{Q}=0.042$ at nucleosynthesis.  Similar  values are found
for Model  II, namely $h=0.6$,  $\Omega_{Q}=0.7$, $\Omega_{m}=0.3$,
$w_{Q}=-0.9$    and   $q=-0.4$    today,    $\Omega_{Q}=0.042$ at
nucleosynthesis. 

Our models seem to be
more sensitive to  changes in  the initial  conditions than models with just 
one scalar field and, in particular,  the  AS model (this is to be expected
since there is more freedom e.g. in the way kinetic energy is shared 
between the two fields) but no fine tuning of the initial conditions 
is needed. Indeed, fixing e.g. 
$x=z$, corresponding to equipartition of kinetic energy
between $\phi$ and $\psi$, we have studied the behaviour of $\rho_Q$  and
seen  that, for a wide range of the remaining initial conditions, after some
initial
transient, each solution scales with the dominant matter component until
$\rho_Q$ begins to dominate.

We have studied the nature of our solutions for a rather broad range
of parameters of the potential. 
We have found that it is possible to
obtain permanent or transient vacuum domination, satisfying present
bounds on observable cosmological parameters, for various combinations 
of the potential parameters.



A  relevant  issue of  our  proposal is  that  it  allows evading  the
conclusions of  Refs. \cite{Hellerman}, in  what concerns the
stability of a supersymmetric  potential.  The main argument presented
in   \cite{Hellerman}  relies   on  the   fact  that,   in  a
supersymmetric theory, one expects  that the asymptotic behaviour of
the superpotential  is given by  $W(\phi) = W_0 e^{-\alpha  \phi /2}$,
which,  in  order to  ensure  the  positivity  of the  $4$-dimensional
potential  $V(\phi) =  8 \vert  \partial_{\phi}  W \vert^2  - 12
  \vert W^2 \vert$
implies that $\vert  \alpha \vert > \sqrt{6}$. However,  this value is
inconsistent  with  the  requirement  of an  accelerated  Universe  at
present  $\vert  \alpha  \vert = \sqrt{3(1+\omega_{Q0})/2} < 1.5$
\cite{Ratra}   as   data   suggest   that  $\omega_{Q0} < - 0.6$
\cite{Perlmutter1}.   The situation  is different  in the  presence of
fields that do not reach  their minima asymptotically, as in Model II.
Indeed, in  this case, the asymptotic behaviour  of the superpotential
would be  better described by  the function $W(\phi) =  W_0 e^{-\alpha
\phi /2} F(\phi, \psi)$, where  $F(\phi, \psi)$ is a polynomial in the
fields  $\phi$  and  $\psi$.    The  positivity  condition  now  reads:
$\alpha^2 - 6 +  4 [(\partial_{\phi} F)^2 + (\partial_{\psi} F)^2]/F^2
> 0$.  One can  then easily  see  that, by  a suitable  choice of  the
polynomial $F(\phi, \psi)$, the positivity condition can be reconciled
with the requirement of successful quintessence. Furthermore, since in
Model II  acceleration is transient  and occurs only at  present, this
model is consistent with the  underlying framework of string theory as
is  does not present  cosmological horizons  that are  associated with
eternally accelerating universes.

Solutions corresponding to transient acceleration have not been found
in previous two-field quintessence models. In the model of
Ref.~\cite{Masiero}, where the fields invoked are the vacuum expectation
values of SUSY QCD chiral superfields, quintessence energy density grows
with respect to matter as $\rho_Q/\rho_m \sim a^{3(1+r)/2}$, where $r$ is
the ratio between the number of flavours and the number of colours. Similarly,
 in the model of Ref.~\cite{Fujii},
where the potential $V(\sigma, \Phi)=e^{-4 \zeta \sigma}
(\Lambda + \frac{1}{2} m^2 \Phi^2 [1+\gamma \sin(k \sigma)])$ is proposed, 
quintessence energy density  dominates matter energy density 
asymptotically.

We conclude that the late time dynamics arising from our two-field
potential is consistent with the observations as well as the
theoretical requirements of stability of the supersymmetric ground
state and  the asymptotic behaviour of string theory states, provided
the observed  accelerated expansion of the Universe is transient and
decelerated expansion is soon resumed, as in Model II.
This solution has been recently
proposed  to solve the  contradiction between  
accelerated expansion and string  theory \cite{Kolda}, on  general
grounds; in this work, we have presented a concrete example of a 
two-field model that exhibits this desirable feature.

M.C.B. thanks T. Barreiro for helpful discussions. M.C.B.  and  O.B.
 acknowledge the partial  support of FCT (Portugal)
under the grant POCTI/1999/FIS/36285.


\end{document}